\documentclass[twocolumn,showpacs,preprintnumbers,amsmath,amssymb]{revtex4}

\usepackage{graphicx}
\usepackage{dcolumn}
\usepackage{bm}
\usepackage{subfigure}
\usepackage{color}

\begin{document}
\title{Fluctuation-Dissipation Relations for Motions of Center of Mass in Driven Granular Fluids under Gravity}

\author{Jun'ichi Wakou$^{1,2}$}
 \email{wakou@cc.miyakonojo-nct.ac.jp}

\author{Masaharu Isobe$^{3}$}
 \email{isobe@nitech.ac.jp}

\affiliation{
$^1$Miyakonojo National College of Technology, Miyakonojo-shi, Miyazaki, 885-8567, Japan\\
$^2$Department of Physics, Kyushu University 33, Fukuoka 812-8581, Japan\\
$^3$Graduate School of Engineering, Nagoya Institute of Technology, 466-8555, Japan
}

\date{\today}

\begin{abstract}
We investigated the validity of fluctuation-dissipation relations in the non-equilibrium stationary state of fluidized granular media under gravity by two independent approaches, based on theory and numerical simulations. A phenomenological Langevin-type theory describing the fluctuation of center of mass height, which was originally constructed for a one-dimensional granular gas on a vibrating bottom plate, was generalized to any dimensionality, even for the case in which the vibrating bottom plate is replaced by a thermal wall. The theory predicts a fluctuation-dissipation relation known to be satisfied at equilibrium, with a modification that replaces the equilibrium temperature by an effective temperature defined by the center of mass kinetic energy. To test the validity of the fluctuation-dissipation relation, we performed extensive and accurate event-driven molecular dynamics simulations for the model system with a thermal wall at the bottom. The power spectrum and response function of the center of
mass height were measured and closely compared with theoretical predictions. It is shown that the fluctuation-dissipation relation for the granular system is satisfied, especially in the high-frequency (short time) region, for a wide range of system parameters. Finally, we describe the relationship between systematic deviations in the low-frequency (long time) region and the time scales of the driven granular system.
\end{abstract}


\maketitle

\section{Introduction}
Granular materials show fluid-like behavior when they are supplied sufficient energy by external vibration. Fluidized states of granular matter have been studied as interesting examples of non-equilibrium fluids. They exhibit a rich variety of phenomena such as convection, pattern formation on the surface, and segregation (see Ref.~\cite{AransonTsimring-2006} and references therein). Besides these pattern-forming instabilities, the plain stationary state of vibrated granular fluids without complex spatial structures serves as an archetypal example of non-equilibrium stationary states (NESSs). It has been a fundamental goal for many years to find any thermodynamic-like description or to identify the common property of fluctuations in a wide variety of NESSs in nature.

One of the important issues addressed in this paper is the validity of the fluctuation-dissipation relation (FDR) in granular fluids subject to external vibration. An FDR connects the response of an equilibrium system to a small perturbation with the time correlation of spontaneous fluctuations in the system without perturbation. Recently, there has been much interest in how an FDR is violated or should be modified in ageing systems, such as glass, and in NESSs of various systems. (See Refs.~\cite{Cugliandolo-2011,MarconiPuglisiRondoniVulpani-2008} for recent reviews.)

For granular systems, FDRs have been studied for several situations. Much work has been devoted to the case of freely cooling granular gas, where the gas develops freely without external forces and ``cools'' as a result of the dissipative nature of the grain interactions. The aim has been to derive a (modified) Green-Kubo relation from which transport coefficients can be calculated~\cite{GoldhirschVanNoije-2000,DuftyGarzo-2001,DuftyBrey-2002,Dufty-BaskaranBrey-2006}. While there is no stationary state for a freely cooling granular gas, a granular gas NESS can be achieved by supplying energy from outside by means of external forcing. A typical experimental means of injecting energy is to shake a container or vibrate a bottom wall (see, e.g., Ref.~\cite{Duran}). In the case where the shaking (or vibrating) is strong enough to inject energy to all grains by frequent collisions with the vibrating wall, the effect of the vibrating wall is often modeled using a thermal bath that couples to every particle. FDRs in
such uniformly driven granular systems have been studied in Refs.~\cite{PuglisiBaldassarriLoreto-2002,Garzo-2004,PuglisiBldassaryVulpiani-2007,BuninShokefLevine-2008,BreyRuiz-Montero-2010}. Puglisi {\it et al}.~\cite{PuglisiBaldassarriLoreto-2002} carried out numerical simulations of a model of uniformly driven granular gas and studied FDRs for two different observables. They observed that the FDRs were satisfied if the equilibrium temperature in the FDR for a system at equilibrium was replaced by the granular temperature, defined as the mean-square fluctuation of the grain velocity. Garz\'o~\cite{Garzo-2004} studied the diffusion of impurities immersed in a granular gas under the influence of uniform driving forces analytically. They showed that a modified form of the Einstein relation, in which the temperature of the gas is replaced by the temperature of the impurity, is violated due to the non-Maxwellian behavior of the impurity velocity distribution function. Bunin {\it et al}.~\cite{BuninShokefLevine-2008} analyzed
a mean-field model of uniformly driven granular gas and showed that the effective temperature defined by an FDR depends on the frequency. In the case where the shaking (or vibrating) is not strong enough to be regarded as uniform driving, energy injection through a boundary has to be explicitly considered. Brey {\it et al}.~\cite{BreyRuiz-Montero-2010} studied the volume fluctuations of a vibrated low-density granular gas confined at the top by a mobile piston numerically. In this system, energy is supplied from the vibrating bottom wall. They discussed the interpretation of an effective temperature defined by requiring the same relation between fluctuations of volume and compressibility as in equilibrium systems. The FDRs and effective temperatures in much denser systems have also been studied by several authors~\cite{MakseKurchan-2002,BarratColizzaLoreto-2002,OhernLiuNagel-2004, DannaMayorBarratLoretoNori-2003}. Among these studies, we refer to an experimental study by D'Anna {\it et al}.~\cite{
DannaMayorBarratLoretoNori-2003}, because their theory based on a Langevin equation formally has the same form as ours, although their experimental setup was very different. They performed an experiment to observe the fluctuating motion of a torsion oscillator immersed in vibration-fluidized granular matter and found that it can be described to first approximation by the formalism for Brownian motion in equilibrium, and an FDR with an effective temperature approximately holds.

We investigated the fluctuating motion of the center of mass (COM) in an NESS of granular matter fluidized by an external energy source located at a bottom wall, under the influence of gravity. Instead of using macroscopic probes such as a piston~\cite{BreyRuiz-Montero-2010} or torsion oscillator~\cite{DannaMayorBarratLoretoNori-2003}, we focused on the position of the COM, which is observable using digital high-speed photography in experiments~\cite{WarrHuntleyJacques-1995}. Our major motivation for studying fluctuations of the COM is that a simple (or universal) law might hold as a result of the following properties. First, the fluctuations of macrovariables such as the COM position often possess the largest time scales in the system. Second, they are expected to be Gaussian in a similar sense to the central limit theorem. (In the case of a Markovian stochastic process, the Gaussian property of macrovariables fluctuations can indeed be derived from a master equation of the Markovian process~\cite{vanKampen}
.)
With this expectation, we proposed a phenomenological theory based on a simple formalism for Brownian motion that describes the motion of the COM height in the NESS of a one-dimensional vibrated granular fluid~\cite{WakouOchiaiIsobe-2008}. We found that the important qualitative features of the dynamics of the COM in event-driven molecular dynamics (MD) simulations were all accounted for by the theory. The theory was extended to a two-dimensional granular fluid on a thermal wall~\cite{WakouIsobe-2010}. Here, we show that when we apply the phenomenological theory to granular fluids in higher dimensions, careful consideration of time scales in granular hydrodynamics~\cite{BreyRuiz-MonteroMoreno-2001,BrombergLivneMeerson-2003} is necessary. Within the time range for which our theory is valid, it predicts the existence of an FDR. However, the equilibrium temperature in the FDR for an equilibrium system must be modified by the effective temperature of the COM velocity fluctuation. To test our prediction, we
performed extensive and accurate event-driven MD simulations for a two-dimensional system of inelastic hard disks on a thermal wall.

Our main result is that an FDR with an effective temperature holds within statistical uncertainty for simulations in a high-frequency (short time) region, while it is violated in a low-frequency (long time) region. The effective temperature is defined by the COM kinetic energy. We observed in our simulations that the ratio between the effective temperature and the global granular temperature increases with inelasticity; the former can be more than four times larger than the latter for the highest inelasticity case.

This paper is organized as follows. In Sec.~II, we describe a model granular system and discuss important time scales in the system. In Sec.~III, the Langevin equation is introduced, and analytical expressions for the power spectrum and response function of the COM height are described briefly. We also remark on the FDR between these two functions. The complete derivation of the Langevin equation and detailed calculation for the power spectrum and the response function are summarized in Appendix A and B, respectively. A comparison between the theoretical predictions and an extensive event-driven MD are shown in Sec.~IV. Finally, in Sec.~V, we summarize the main results for FDR validity and comment on the relation between the systematic deviations in the low-frequency (long time) region and the time scales of the driven granular system.

\section{The model system}
\subsection{System}
As a model of grains bouncing on a vibrating bottom plate under gravity, we consider a $d$-dimensional system of $N$ inelastic particles on a ``thermal'' bottom wall in a constant gravitational field $g$. The particles in the system have diameter $\sigma$ and mass $m$; the total mass of particles is denoted by $M$ ($=Nm$). The thermal wall is kept at a constant temperature $T_0$, which plays the role of a heat source supplying sufficient translational energy to the particles to fluidize them. The $z$-direction is chosen to be opposite to the direction of gravity, and the thermal wall is fixed at $z=0$. For simplicity, we adopt periodic boundary conditions in horizontal directions, so as to ignore the boundary (side-wall) effects. Collisions between particles are inelastic; inelasticity of the particle collisions is characterized by a normal restitution coefficient $r$. To avoid any pattern-forming instability in the horizontal directions, we chose both the inelasticity and linear scales of the system in
horizontal directions to be sufficiently small that the system remained homogeneous in the horizontal directions. These conditions are discussed in more detail in Sec.~IV.

\subsection{Time scales}
Before discussing the important time scales in the system, we define several quantities that characterize the macroscopic properties of the system. We first define the kinetic energy per particle as $K(t)\equiv (1/N)\sum_{i=1}^{N} mv_i(t)^2/2$ and the long time average of $K(t)$ in an NESS as $\overline{K}\equiv \lim_{T\to\infty}(1/T)\int_{0}^{T}K(t)dt$ (hereafter, the overline on a quantity represents its long time average in an NESS). We also define the global granular temperature $T$ as $k_B T\equiv (2/d)\overline{K}$, where $k_B$ is the Boltzmann constant, and the thermal velocity as $c\equiv (dk_B T/m)^{1/2}=(2\overline{K}/m)^{1/2}$. A characteristic length scale of the system in the vertical direction $l$ is then defined as $l\equiv c^2/g$.

Bromberg {\it et al}.~\cite{BrombergLivneMeerson-2003} have suggested that there are three important time scales in this system at the hydrodynamical level: the macroscopic oscillation time $\tau_{\rm osc}$ (referred to as the ``fast time scale'' in Ref.~\cite{BrombergLivneMeerson-2003}), the relaxation time for thermal conduction $\tau_{\rm therm}$, and the relaxation time for collisional dissipation $\tau_{\rm diss}$.

For simplicity, we assume that the system is nearly homogeneous, although this is not true for small $r$ and large $N$. The time scale $\tau_{\rm osc}$ represents the period of the slowest oscillation in the vertical direction, that is, the period of the sound mode with the longest wavelength. Thus, $\tau_{\rm osc}\sim l/c_s$, where $c_s$ is the sound velocity. Assuming $c_s\sim c$, which is satisfied for a normal gas, $\tau_{\rm osc}$ can be estimated as $\tau_{\rm osc}\sim c/g$. Because $l/c_s$ also characterizes the pressure relaxation time $\tau_p$, we can regard $\tau_p$ and $\tau_{\rm osc}$ as on the same order, $\tau_p\sim\tau_{\rm osc}\sim c/g$. The relaxation time for thermal conduction $\tau_{\rm therm}$ is estimated as $\tau_{\rm therm}\sim l^2/(\kappa/\rho c_p)$, where $\kappa$ is the thermal conductivity, $\rho$ is the mass density, and $c_p$ is the specific heat at constant pressure~\cite{LandauLifshitzFluid}. $\rho$ can be estimated as $\rho\sim M/(l A)\sim mN_z/(l \sigma^{d-1})$, where $A$ represents the area of the bottom
plate in three dimensions ($A$ represents the length of the bottom plate in two dimensions and $A=1$ in one dimension) and $N_z$ represents the number of monolayers at rest. $\kappa$ and $c_p$ are obtained from kinetic theory for elastic spheres and disks~\cite{ChapmanCawling}: $\kappa\sim k_B c/\sigma^{d-1}$ and $c_p\sim k_B/m$. Substituting these results, we obtain $\tau_{\rm therm}\sim N_z c/g$. The relaxation time for collisional dissipation $\tau_{\rm diss}$ can be estimated as the inverse of $(1-r^2)\nu$, where $\nu$ is the collision frequency between two particles. Substituting the lowest order estimation of $\nu$ based on kinetic theory, $\nu\sim \rho \sigma^{d-1} c/m\sim N_z c /l$, we obtain $\tau_{\rm diss}\sim (1-r^2)^{-1}c / (N_z g)$.

The time scales estimated above are summarized as follows:
\begin{eqnarray}
&& \hspace{-0.8cm}
\tau_{\rm osc} \sim \tau_{p}\sim \frac{c}{g},
\hspace{0.5cm}
 \tau_{\rm therm}\sim N_z \frac{c}{g},
\nonumber\\
&& \hspace{-0.8cm}
\tau_{\rm diss} \sim \left[N_z(1-r^2)\right]^{-1}\frac{c}{g}.
\label{timescales}
\end{eqnarray}
It is important to note that all time scales, $\tau_{\rm osc}$, $\tau_{p}$, $\tau_{\rm therm}$, and $\tau_{\rm diss}$, are proportional to $c/g$. This means that for a system with given $N_z$ and $r$, the macroscopic dynamics with time scaled by $c/g$ are independent of $g$. We utilize this fact later to obtain a frequency response function in an efficient way.

There are three dimensionless parameters, obtained as the ratios between two of these three time scales. The first is $\tau_{\rm therm}/\tau_{\rm osc}\sim N_z$. The second is $\tau_{\rm osc}/\tau_{\rm diss}\sim N_z(1-r^2)$. The third is $\tau_{\rm therm}/\tau_{\rm diss}\sim N_z^2(1-r^2)$. The first and third parameters are the governing parameters for the hydrodynamic description of the system, as introduced by Bromberg {\it et al}.~\cite{BrombergLivneMeerson-2003}. They showed that the steady-state profile is governed only by the parameter
\begin{eqnarray}
\Lambda\equiv \frac{\sqrt{\pi}}{2}N_z(1-r^2)^{1/2},
\label{lambda}
\end{eqnarray}
which is proportional to $(\tau_{\rm therm}/\tau_{\rm diss})^{1/2}$. If $1-r\ll 1$, the second parameter $\tau_{\rm osc}/\tau_{\rm diss}$ is related to $X\equiv N_z(1-r)$. It plays the role of the governing transition parameter from a condensed to fluidized state in a one-dimensional column of beads on a vibrating bottom plate~\cite{LudingClementBlumenRajchenbachDuran-1994}. In our study, we consider the case $N_z\gg 1$ and assume $\tau_{\rm therm}\gg \tau_{\rm osc}$ in the following theoretical analysis.

\section{Theoretical Derivation of the Fluctuation-dissipation Relation}

Here, we summarize the theoretical derivation of (i) the power spectrum, (ii) the frequency response function, and (iii) the FDR between (i) and (ii). First, we introduce a Langevin equation as a first approximation that describes the fluctuating motion of the COM on the fast time scales $\tau_{\rm osc}$ and $\tau_p$. Note that the derivation of our theory has already been published in Ref.~\cite{WakouOchiaiIsobe-2008}. We assume $\tau_{\rm therm}\gg\tau_{\rm osc}$, as mentioned above, and focus on the dynamics of the COM on the time scale $\tau_{\rm osc}$, ignoring the significant slow relaxation process of fluctuations of global granular temperature around its stationary value $(2/d)\overline{K}/k_{B}$. The effect of this slow dynamics of granular temperature and validity of our time scale assumption are discussed later.

We summarize the details of the derivation of our Langevin formalism in Appendix A and show the final result here. We denote the height of the COM of granular fluids at time $t$ as $Z(t)$, the time average of $Z(t)$ over a long time interval in an NESS as $\overline{Z}$, and small deviations of $Z(t)$ from $\overline{Z}$ as $\delta Z(t)\equiv Z(t)-\overline{Z}$. The Langevin equation for fluctuating motion of $\delta Z(t)$ is given by (see Eq.~(\ref{langevin1}) in Appendix A)
\begin{eqnarray}
\frac{d^2\delta Z}{dt^2}=-\Omega^2\delta{Z}-\mu\frac{d \delta Z}{dt} + \frac{R(t)}{M},
\label{langevin}
\end{eqnarray}
where $R(t)$ represents a random force, which is assumed to be a Gaussian white noise:
\begin{equation}
\left<R(t)\right>=0,\quad \quad \left<R(t)R(t')\right>=I\delta(t-t').
\label{gaussianwhite}
\end{equation}
The brackets $\left<\cdots\right>$ denote an average over the random force. In NESS, it is reasonable to assume $\left<Z(t)\right>_{\rm st}=\overline{Z}$, where $\left<\cdots\right>_{\rm st}$ represents the average in a stationary state. The constant $I$ represents the intensity of the random force, which is related to the second moment of the COM velocity fluctuations. This relation can be obtained by calculating the average kinetic energy of the COM motion in $z$-direction
$K_{\rm COMz}\equiv \left< M V_z(t)^2/2\right>_{\rm st}$, where $V_z$ is the $z$-component of the velocity of the COM, $V_z(t)\equiv \frac{d Z(t)}{dt}$. Using an analytical solution Eq.~(\ref{cmheight}) of the Langevin equation, we obtain $K_{\rm COMz}=I/4M\mu$. Hence, the constant $I$ is identified as
\begin{eqnarray}
I\equiv 4M\mu K_{\rm COMz}.
\label{intensity}
\end{eqnarray}

This is the same procedure used to determine the noise intensity $I$ when the Langevin equation describes fluctuations in equilibrium at temperature $T$. In equilibrium, equipartition of energy implies $K_{\rm COMz}=k_B T/2$, that is, the mean kinetic energy of the COM in the $z$-direction $K_{\rm COMz}$ and the mean kinetic energy of a particle in one direction $k_B T/2$ are the same. Thus, we obtain the well-known result $I = 2M\mu k_B T$. In the case of an NESS of granular fluids, the violation of equipartition of energy is observed in various systems. A heated binary granular system (see Ref.~\cite{WangMenon-2008} and references therein) is one notable example in which non-equipartition between the mean kinetic energies of two species has been studied. Later, we present numerical simulations that clearly show violation of equipartition, $K_{\rm COMz}\neq k_B T/2$, when we recognize $T$ as the global granular temperature.

The coefficients $\Omega$ and $\mu$ describe an angular frequency of the slowest oscillation of the COM height and frictional coefficient with respect to relative motion of the COM height against the bottom wall, respectively. According to the time scales we consider here, we assume $\Omega\sim \tau_{\rm osc}^{-1}$ and $\mu \sim \tau_{p}^{-1}$ and write them as
\begin{eqnarray}
\Omega=\hat{\Omega}g/c,\hspace{0.5cm}
\mu=\hat{\mu} g/c.
\label{coefficients}
\end{eqnarray}
Because values of the coefficients $\hat{\Omega}$ and $\hat{\mu}$ cannot be estimated in our phenomenological theory, they are fixed as fitting parameters when we compare results of simulations with the theoretical predictions.

{\bf Power Spectrum} The power spectrum $S(\omega)$ that represents the fluctuations of $Z$ around the NESS is defined as the Fourier transform of the time correlation function,
\begin{eqnarray}
S(\omega)\equiv \int_{-\infty}^{\infty}dt \,e^{-i\omega t}\left<\delta Z(0)\delta Z(t)\right>_{\rm st}.
\label{psdefinition}
\end{eqnarray}
The derivation of $S(\omega)$ using the analytic solution of the Langevin equation is straightforward. The final expression of $S(\omega)$ in this system is
\begin{eqnarray}
S(\omega)
&=&
\frac{1}{M}\frac{4\mu K_{\rm COMz}}{\left(\Omega^2-\omega^2\right)^2+\left(\mu \omega\right)^2}.
\label{powerspectrum}
\end{eqnarray}
See Appendix B for a detailed derivation.

{\bf Response Function}
The frequency response function $\chi(\omega)$ that characterizes the linear response of $Z$ in the NESS against a small external force $\varepsilon f(t)$ can be defined as
\begin{eqnarray}
 \chi(\omega)\equiv \lim_{\varepsilon\to 0} \left<\delta \tilde{Z}(\omega)\right> / \varepsilon \tilde{f}(\omega),
\label{resdefinition}
\end{eqnarray}
where $\delta \tilde{Z}(\omega)$ and $\tilde{f}(\omega)$ are the Fourier transform of $\delta Z(t)$ and $f(t)$, respectively. The analytical expression of $\chi(\omega)$ is given as
\begin{eqnarray}
\chi(\omega)=\frac{1}{M}\frac{1}{\Omega^2-\omega^2+i\mu\omega}.
\label{chi}
\end{eqnarray}
 A detailed derivation is given in Appendix B.
According to conventional definition, $\chi(\omega)$ can be decomposed into real $\chi'(\omega)$ and imaginary $\chi''(\omega)$ parts as $\chi(\omega)=\chi'(\omega)-i\chi''(\omega)$. Thus, we obtain the expression
\begin{eqnarray}
\chi'(\omega)&=&\frac{1}{M}\frac{\Omega^2-\omega^2}{(\Omega^2-\omega^2)^2+(\mu\omega)^2},
\label{chi'}
\end{eqnarray}
\begin{eqnarray}
\chi''(\omega)&=&\frac{1}{M}\frac{\mu\omega}{(\Omega^2-\omega^2)^2+(\mu\omega)^2}.
\label{chi"}
\end{eqnarray}

{\bf Fluctuation Dissipation Relation}
Comparing Eqs.~(\ref{powerspectrum}) and (\ref{chi"}), we obtain the FDR
\begin{eqnarray}
\frac{\omega S(\omega)}{2k_B T_{eff}}=\chi''(\omega),
\label{fdt}
\end{eqnarray}
where $T_{eff}$ is an effective temperature defined as $T_{eff}\equiv 2K_{\rm COMz}/k_B$. This has the same form as the FDR in an equilibrium system except for $T_{eff}$, which replaces the equilibrium temperature.

\section{Numerical simulations}
Here, we compare the three theoretical predictions described in the previous section with results of the numerical simulation of a two-dimensional granular gas system. The predictions are the power spectrum Eq.~(\ref{powerspectrum}), the frequency response function Eq.~(\ref{chi}), and the fluctuation-dissipation relation Eq.~(\ref{fdt}). Our system consisted of $N$ inelastic hard disks of mass $m$ and diameter $\sigma$ moving in two dimensions on a thermal wall with a fixed temperature $T_0$. Here, the $x$- and $z$-axes represent the horizontal and vertical directions of the system, respectively. The system width is denoted as $L$, and periodic boundary conditions were adopted in the horizontal direction at $x=0$ and $x=L$. The bottom wall was located at $z=0$, and there was no top wall. Gravitational force was exerted on each disk along the negative $z$-direction. Inelastic collisions between hard disks were considered by the normal restitution coefficient $r$. When a disk collided with a thermal wall at
the bottom, it left with a value of $z$-component of velocity $v_z$ sampled from the probability density
\begin{eqnarray}
p(v_z)=\frac{mv_z}{k_B T_0}\exp \left(-\frac{mv_z^2}{2k_B T_0}\right).
\end{eqnarray}
The horizontal component of velocity did not change during the collision.

Numerical simulations were performed with an event-driven algorithm devised to enhance the speed of calculation in dense hard sphere systems~\cite{Isobe-1999}. In the following, all simulation data are presented with mass, length, and time in units of $m$, $\sigma$, and $\sigma/(k_B T_0/m)^{1/2}$, respectively. This corresponds to choosing $k_B T_0=1$. We set $N=5000$ and $L=100$ (these parameters are unchanged throughout this paper). For our main results, $r=0.99- 0.999$ and $g =10^{-3}$ were used unless otherwise mentioned. These correspond to $N_z=50$, $0.05\le X \le 0.5$ and $1.98\le \Lambda<6.25$. A system of width $L=100$ for  $r\ge 0.99$ is small enough to prevent any horizontal pattern formation (e.g., ripples or undulations). The global temperature $T$ and the thermal velocity $c$ were calculated using $T=\overline{K}/k_{B}$ and $c=(2\overline{K}/m)^{1/2}$, where $\overline{K}$ is the long time average of  the kinetic energy per disk.

\subsection{Macroscopic properties in the NESS}
\begin{figure}[htbp]
\begin{center}
\subfigure{%
 \includegraphics[width=0.65\columnwidth,clip]{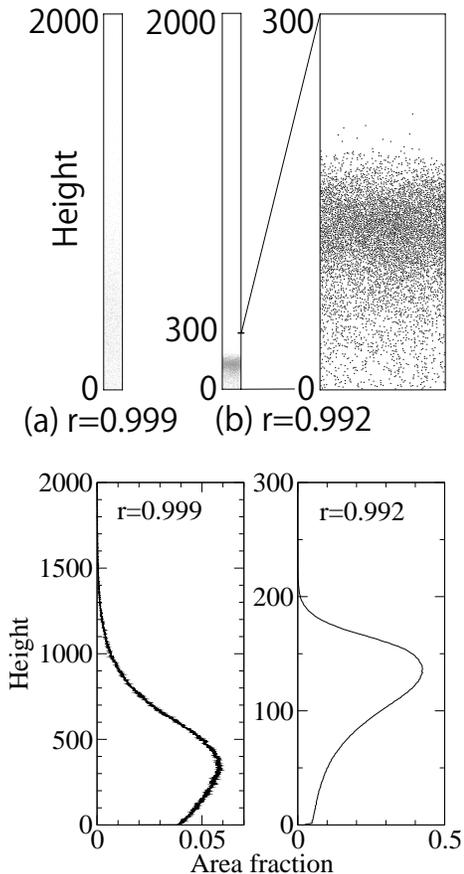}}%

\subfigure{%
 \includegraphics[width=0.7\columnwidth,clip]{./fig1_bottom.eps}}%
\caption{\label{snapshot}
Top: Snapshots of the two-dimensional simulation with $N=5000$, $L=100$, and $g=10^{-3}$ for different values of the restitution coefficient (a) $r=0.999$; (b) $r=0.992$. Bottom: Area-fraction profiles averaged over a long time period for $r=0.999$ and $r=0.992$.
}
\end{center}
\end{figure}

In Fig.~\ref{snapshot} (top), we show typical snapshots of particle configurations in the system of $N=5000$ and $g=10^{-3}$ for $r=0.999$ and $0.992$. The corresponding area-fraction profiles are plotted in Fig.~\ref{snapshot} (bottom). For a nearly elastic case ($r=0.999$), the profile had one peak around the height $z\simeq 350$. However, the area fraction was relatively dilute (less than $~0.06$), even at the height of the peak. Many inelastic particles were raised up relatively high, like the equilibrium profile of the Boltzmann distribution. In contrast, for $r=0.992$, the profile drastically changed. Most particles condensed at a relatively low level in a cluster; the area-fraction profile showed a clear peak above the low-density region around the thermal wall. This state is known as {\it density inversion state} and has been observed in many experiments~\cite{KudrolliWolpertGollub-1997,WildmanHuntleyHansen-2001} and simulations~\cite{LanRosato-1995,IsobeNakanishi-1999} of vibrofluidized granular
matter.

\begin{figure}[htbp]
\begin{center}
 \includegraphics[width=0.8\columnwidth,clip]{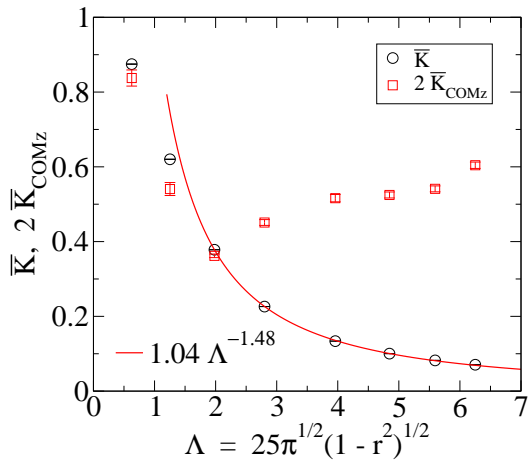}
\caption{\label{kineticenergies} (Color online)
Kinetic energy per particle $\overline{K}$ (circles) and kinetic energy of the COM $\overline{K}_{\rm COMz}$ (squares), plotted versus $\Lambda=\frac{\sqrt{\pi}}{2}50(1-r^2)^{1/2}$ for $r=0.9999,\,0.9996,\,0.999,\,0.998,\,0.996,\,0.994,\,0.992$, and $0.99$ from left to right. The solid line gives a numerical fit of the form $1.04\times\Lambda^{-1.48}$.
}
\end{center}
\end{figure}

In accordance with the theoretical study by Bromberg {\it et al}.~\cite{BrombergLivneMeerson-2003}, which showed that the steady state is characterized by a single parameter $\Lambda$, defined in Eq.~(\ref{lambda}), we plotted the average kinetic energy per disk $\overline{K}$ and the average kinetic energy of the COM $\overline{K}_{\rm COMz}$ as a function of $\Lambda$ in Fig.~\ref{kineticenergies}. The statistical error bars with standard deviation were also plotted in all figures throughout the paper. $\Lambda=0$ (that is, $r=1$) corresponds to the equilibrium state in which equipartition of energy $2\overline{K}_{\rm COMz}=\overline{K}=k_B T_0=1$ is satisfied. The factor $2$ comes from the fact that $\overline{K}_{\rm COMz}$ is defined using only the $z$-component of the COM velocity. The horizontal component of the velocity of the COM vanished in our simulations because the horizontal component of disk velocity was unchanged on collision with the bottom wall. While $\overline{K}$ systematically decreased following a power
law $\sim \Lambda^{-1.48}$, $\overline{K}_{\rm COMz}$ reached a minimum at $\Lambda\simeq 2$ ($r=0.999$) and increased with $\Lambda$ for $\Lambda > 2$. Fig.~\ref{kineticenergies} clearly indicates that equipartition of energy breaks down when $\Lambda > 2$ (that is, $2\overline{K}_{\rm COMz}\ne \overline{K}$). Similar behavior in much smaller systems has been reported by us~\cite{WakouIsobe-2010}. In Ref.~\cite{BrombergLivneMeerson-2003}, it was shown that the density inversion appears above the threshold $\Lambda_c$ ($\Lambda>\Lambda_c$), where $\Lambda_c\simeq 1.06569$. In the density inversion state, which becomes pronounced for $\Lambda>2$, as shown in Fig.~\ref{snapshot}, a low-density and high-temperature gaseous region near the bottom can cause large fluctuations of the dense cluster on top. Therefore, this violation of the equipartition of energy should be closely connected to development of the density inversion.

The relation between the long time average of the COM height $\overline{Z}$ and the kinetic energy per particle is given in Fig.~3. A linear relation $\overline{Z}=\overline{K}/mg+\mbox{const.}$ was satisfied for $\overline{K}>0.1$, even when the system had a density inversion with a relatively high density cluster. In an equilibrium system of dilute gases, the relation $\overline{Z}=\overline{K}/mg+\mbox{const.}$ holds as a result of statistical mechanics. The fact that $\overline{K}$ characterizes $\overline{Z}$ in the same way as in equilibrium suggests that the global granular temperature $T$ in the \emph{inhomogeneous} non-equilibrium state still retains the same meaning as the equilibrium temperature, at least in a macroscopic sense.
\begin{figure}[htbp]
\begin{center}
 \includegraphics[width=0.8\columnwidth,clip]{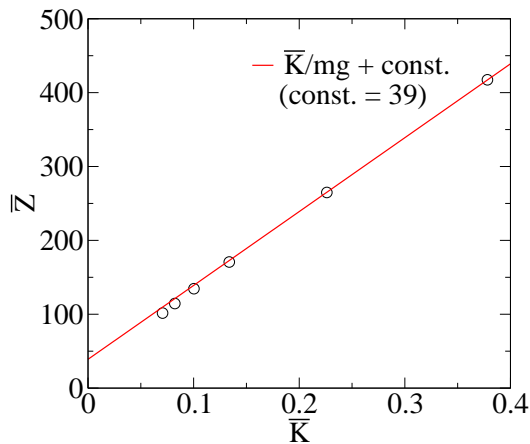}
\caption{\label{kin-pot} (Color online)
The average height of the center of mass $\overline{Z}$ versus $\overline{K}$ for $r=0.999,\,0.998,\,0.996,\,0.994,\,0.992$, and $0.99$ from right to left. The error bars are smaller than the size of the marks. The solid line gives a linear fit with the slope $(mg)^{-1}$, where $m=1$ and $g=10^{-3}$.
}
\end{center}
\end{figure}

\begin{figure}[htbp]
\begin{center}
 \includegraphics[width=0.8\columnwidth,clip]{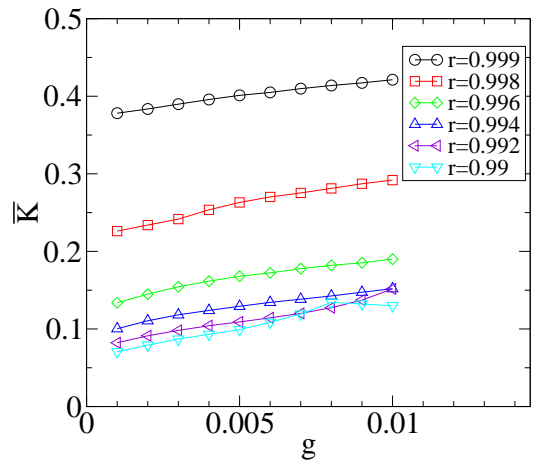}
\caption{\label{kin-G} (Color online)
Kinetic energy per particle $\overline{K}$ as a function of $g$. The error bars are smaller than the sizes of the marks.
}
\end{center}
\end{figure}

In Fig.~\ref{kin-G}, $\overline{K}$ is plotted as a function of the gravitational acceleration $g$. The dependence of $\overline{K}$ on $g$ turned out to be rather weak. We utilized this fact to measure the response function from simulations in an efficient way (see Sec.~IV~C).

\begin{figure}[tbp]
\begin{center}
 \includegraphics[width=0.8\columnwidth,clip]{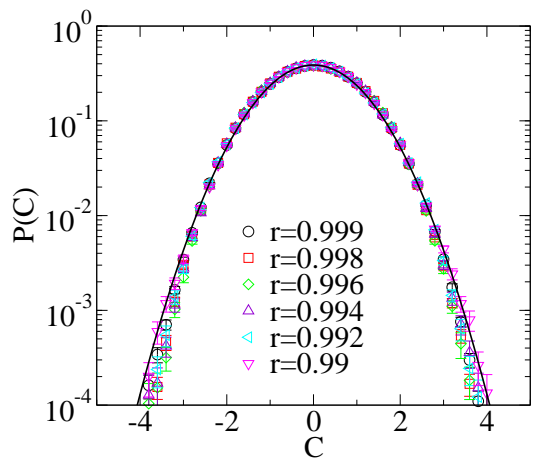}
\caption{\label{vdistrib} (Color online)
Probability distribution of the scaled COM velocity $C\equiv V_z/(2 \overline{K}_{\rm COMz}/M)^{1/2}$. The solid line is Gaussian with unity dispersion.
}
\end{center}
\end{figure}

In Fig.~\ref{vdistrib}, we plotted the probability distribution $P(C)$ of the scaled COM velocity $C\equiv V_z/(2 \overline{K}_{\rm COMz}/M)^{1/2}$. The data were fitted sufficiently by a Gaussian for all cases studied in this paper, as expected from the central limit theorem. This Gaussian property was consistent with our theory based on a linear Langevin equation with additive Gaussian noise.

\subsection{Power spectrum of the COM height}
We first tested the theoretical prediction Eq.~(\ref{powerspectrum}) for the power spectrum of the COM height. Using the relation Eq.~(\ref{coefficients}), Eq.~(\ref{powerspectrum}) can be rewritten as
\begin{eqnarray}
\hat{S}(\hat{\omega})
&\equiv&
S(\hat{\omega} g/c)/\left[ 4\left(\frac{c}{g}\right)^3 \frac{\overline{K}_{\rm COMz}}{M}\right]
\nonumber\\
&=&
\frac{\hat{\mu}}{\left(\hat{\Omega}^2-\hat{\omega}^2\right)^2+\left(\hat{\mu}\hat{\omega}\right)^2},
\label{scps}
\end{eqnarray}
where $\hat{\omega}$ is the scaled angular frequency, defined by $\hat{\omega}\equiv \omega c/g$. This expression suggests that if we scale the power spectrum and the angular frequency as in Eq.~(\ref{scps}), it shows a universal behavior independent of any system parameters.

\begin{figure}[htbp]
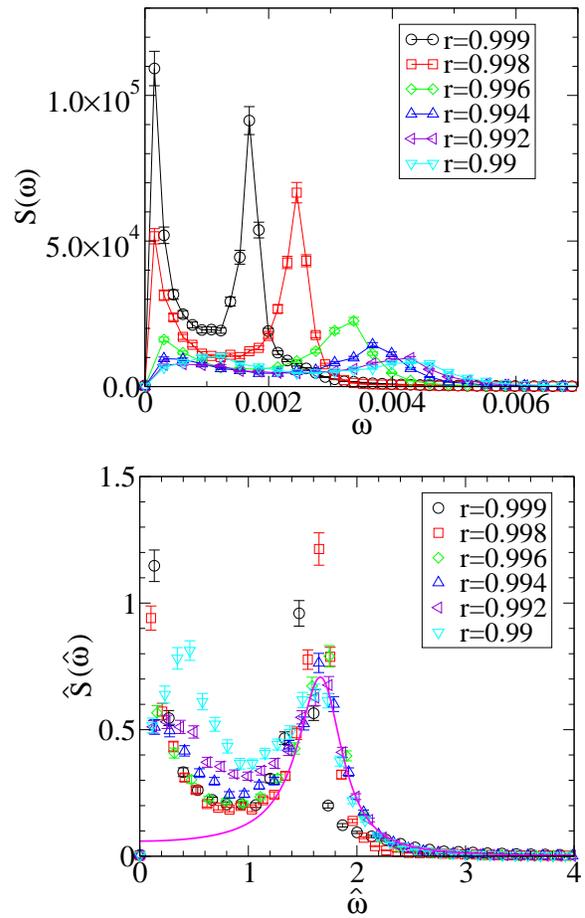

\begin{center}
\hspace{-0.8cm}
\subfigure{%
 \includegraphics[width=0.87\columnwidth,clip]{./fig6_top.eps}}%

\subfigure{%
 \includegraphics[width=0.8\columnwidth,clip]{./fig6_bottom.eps}}%
\caption{\label{pscom} (Color online)
Top: Power spectrum for the COM height versus angular frequency $\omega$. Averages were taken over 400 realizations. Bottom: Scaled power spectrum for the COM height. The solid line depicts the theoretical prediction given in Eq.~(\ref{scps}) with $\hat{\mu}=0.50$ and $\hat{\Omega}=1.7$.
}
\end{center}
\end{figure}

In Fig.~\ref{pscom} (top), the power spectrum $S(\omega)$ is plotted for different values of $r$. Two sharp peaks were observed; one is near zero angular frequency ($\omega=0$), the other one is at the angular frequency of the macroscopic oscillation ($\omega=\omega_{\rm osc}$), which increased as $r$ decreased. The heights of both these peaks decreased with $r$. Figure~\ref{pscom} (bottom) shows the scaled power spectrum $\hat{S}(\hat{\omega})$ obtained by scaling $S(\omega)$ in Fig.~\ref{pscom} (top), according to Eq.~(\ref{scps}) using $c$ and $\overline{K}_{\rm COMz}$ calculated from simulation data. The theoretical prediction Eq.~(\ref{scps}) with fitting numerical parameters $\hat{\mu}=0.50$, $\hat{\Omega}=1.7$ is presented as a thick solid line. It is consistent with the results of simulations for the range $0.99\le r\le 0.996$ in this region near the peak at $\hat{\omega}=\hat{\omega}_{\rm osc}\equiv\omega_{\rm osc}c/g$, where we expect our theory to serve as a first-order approximation. We found large
deviations from the theoretical prediction in the region $\hat{\omega}<\hat{\omega}_{\rm osc}$ (the sharp peak near $\hat{\omega}=0$). As we illustrate below, the peak near $\hat{\omega}=0$ could be associated with slow fluctuations of global granular temperature due to thermal conduction and collisional dissipation. Because $\tau_{\rm therm}/\tau_{\rm osc}\sim N_z\gg 1$ and $\tau_{\rm diss}/\tau_{\rm osc}\sim [N_z(1-r^2)]^{-1}\ge 1.0$ for our simulations with $N_z=50$ and $r\ge 0.99$, the contributions of these two processes should appear at $\hat{\omega}<\hat{\omega}_{\rm osc}$. We also found that for $r\ge 0.998$, the simulation data in Fig.~\ref{pscom} (bottom) deviated from our theory, even in the region near the peak at $\hat{\omega}=\hat{\omega}_{\rm osc}$. These deviations near $\hat{\omega}_{\rm osc}$ could be attributed to the drastic change in density profiles shown in Fig.~\ref{snapshot} as $r$ is varied. Concerning our theory, the change in density profiles may affect the numerical coefficients $\hat{\Omega}$ and $\hat{\mu}$
in Eq.~(\ref{coefficients}). Furthermore, in the region $\hat{\omega}<\hat{\omega}_{\rm osc}$, the effect of global temperature fluctuations mentioned above could become pronounced for $r\ge 0.998$, because both $\tau_{\rm therm}$ and $\tau_{\rm diss}$ became much larger than $\tau_{\rm osc}$, and hence the fluctuations had long lifetimes. Nonetheless, a satisfactory explanation of these deviations for $r\ge 0.998$ has not yet been given.

\begin{figure}[htbp]
\begin{center}
\subfigure{%
 \includegraphics[width=0.8\columnwidth,clip]{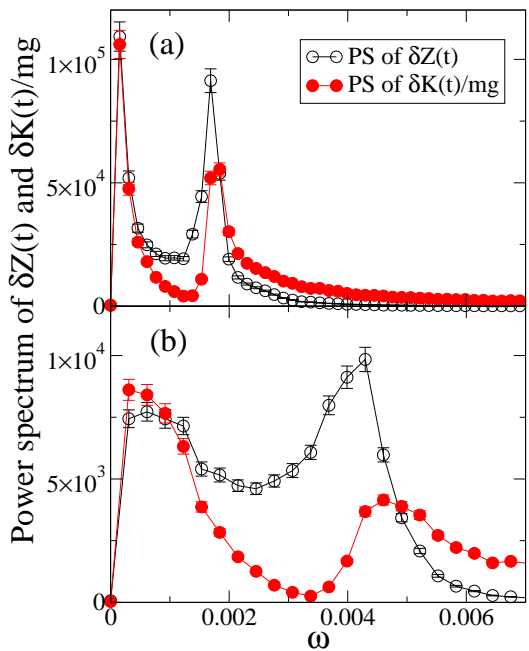}}%
\vspace{-2.7cm}
\caption{\label{pskin} (Color online)
Power spectrum (PS) of $\delta K(t)/mg$ and of $\delta Z(t)$, $S(\omega)$, for (a) $r=0.999$ and (b) $r=0.992$.
}
\end{center}
\end{figure}

Here, we show simulations suggesting that the behavior of $S(\omega)$ in the region near $\omega =0$ can be described by taking into account the slow dynamics of $K(t)$. We denote the slowly varying part of $K(t)$ as $K'(t)$ and suppose it fluctuates on a much longer time scale than $\tau_{\rm osc}$, due to thermal conduction and collisional dissipation. Then, $K'(t)/k_{B}$ can be regarded as a time-dependent global granular temperature. Similarly, we let $Z'(t)$ denote the slowly varying part of $Z(t)$ on the same time scale as $K'(t)$. We assume here that in this long time scale, $K'(t)$ and $Z'(t)$ play the same role as their long time averages $\overline{K}$ and $\overline{Z}$. That is, they satisfy the same linear relation as their long time averages observed in Fig.~\ref{kin-pot}: $Z'(t)=K'(t)/mg+\mbox{const.}$ with the same constant factor. If this is the case, the power spectrum of $\delta Z(t)$, $S(\omega)$, in the region near $\omega =0$ should be given by the power spectrum of $\delta K'(t)/mg$, where
$\delta K'(t)=K'(t)-\overline{K}$. In Fig.~\ref{pskin}, we show the power spectrum of $\delta K(t)/mg$, where  $\delta K(t)=K(t)-\overline{K}$, and $S(\omega)$ for $r=0.999$ and $0.992$. The figure shows that the curves around the peak in $S(\omega)$ near $\omega =0$ and the peak in the power spectrum of $\delta K(t)/mg$ near $\omega =0$ are consistent. The consistency between the two curves is also observed for the other $r$ values. This result indicates that the peak in $S(\omega)$ near $\omega =0$ can be accounted for by slow dynamics of $K(t)$ due to thermal conduction and collisional dissipation. It should be emphasized that in our present theory, fluctuations of granular temperature in both space and time are ignored and only the global granular temperature $T$ is defined, using the long time average of $K(t)$. Further investigation is necessary to construct a theory that fully describes the behavior of $S(\omega)$, taking into account the effect of slow fluctuations of granular temperature.

\subsection{Response functions}
Next, we test the theoretical prediction Eqs.~(\ref{chi'}) and (\ref{chi"}) for the frequency response functions of the COM. By scaling these functions in the same way as the power spectrum, we can derive universal equations
\begin{eqnarray}
\hat{\chi}'(\hat{\omega})
\equiv
\chi'(\hat{\omega} g/c)Mg^2/c^2
=\frac{\hat{\Omega}^2-\hat{\omega}^2}{\left(\hat{\Omega}^2-\hat{\omega}^2\right)^2+\left(\hat{\mu}\hat{\omega}\right)^2},
\label{scaledchireal}
\end{eqnarray}
\begin{eqnarray}
\hat{\chi}''(\hat{\omega})
\equiv
\chi''(\hat{\omega}g/c)Mg^2/c^2
=\frac{\hat{\mu}\hat{\omega}}{\left(\hat{\Omega}^2-\hat{\omega}^2\right)^2+\left(\hat{\mu}\hat{\omega}\right)^2}.
\label{scaledchiimag}
\end{eqnarray}

The frequency response function was measured using numerical simulations via the following procedure. First, we prepared for a system in the stationary state with a given  $N_z$, $r$, and $g$ after a sufficiently long relaxation time from the initial state of particles with randomly distributed positions and velocities. At $t=0$, we exerted a small constant external force on all particles in the direction of gravity and measured the height of the COM at $t>0$; from this COM relaxation process, we deduced a response function by the standard procedure given in textbooks (see, e.g., Ref.~\cite{KuboTodaHashitsume}). In other words, we measured a response function against a step functional external force. The frequency response function was obtained as the Fourier transform of the response function.

It is important to note that in the response of the COM height against a small but finite external force in our system, \emph{nonlinear} effects resulting from time scale changes were non-negligible. This can be seen from the fact that the relevant time scales shown in Eq.~(\ref{timescales}) all depended on $g$ and that exerting a constant force in the direction of gravity was equivalent to changing $g$. Therefore, a linear response could be defined only in the limit of small external force. This shows that our Langevin-type theory is different from the well-known Langevin theory for Brownian motion in a fixed harmonic potential, where the response of a Brownian particle is linear against a finite external force. Consequently, we had to exert an external force that was much smaller than the gravitational force in our system, in order to measure the linear response of the COM height. Because the fluctuation of the COM height of $5000$ particles was typically much larger than the response against such a small
constant force, we needed to perform the response function measurement for a large number of systems with the same parameters $N_z$, $r$, and $g$ but different initial conditions and take an average of the response functions over all realizations. As shown later, in the case of a constant force that is $1\%$ of the gravitational force, we needed more than $10^4$ realizations to obtain sufficient statistics for clear response functions. This required relatively long CPU times that impeded long simulations with a wide range of parameters $r$, $N_z$, and $g$.

We therefore optimized the method by choosing an appropriate parameter to approximately evaluate the response function from a small number of realizations, which could be provided in an acceptable time with our computational facilities. Suppose a system with gravitational field $g$ is initially in an NESS and the gravitational acceleration is increased at $t=0$ from $g$ to $g+\Delta g$. This is equivalent to exerting a step function external force $-M\Delta g\theta(t)$ on the COM height, where $\theta(t)$ is the Heaviside unit step function. Now we define the function $\chi(t;g,g+\Delta g)$ as
\begin{eqnarray}
 \chi(t;g,g+\Delta g)\equiv -\frac{d\left<\delta Z\right>_{t}}{dt}/M\Delta g,
\label{chideltag}
\end{eqnarray}
where $\left<\cdots\right>_{t}$ represents the average taken over the ensemble of realizations $\delta Z$ at time $t$. This is a function of $\Delta g$ in our system due to the nonlinear effects mentioned above; it would equal the response function only if $\left<\delta Z\right>_t$ were linear in $\Delta g$. We denote the Fourier transform of $\chi(t;g,g+\Delta g)$ as $\chi(\omega;g,g+\Delta g)$. According to Eq.~(\ref{resdefinition}), the frequency response function $\chi(\omega;g)$ for the system in the stationary state with $g$ is given by
\begin{eqnarray}
 \chi(\omega;g)=\lim_{\Delta g\to 0}\chi(\omega;g,g+\Delta g).
\label{chiomegag}
\end{eqnarray}

We now consider the time scales that we introduced in Sec.~II~B, which characterize macroscopic dynamics at $t>0$. As we discussed in Sec.~II~B, all these time scales in the NESS depend on $g$ in the form $\tau= c(g)/g\times \mbox{const.}$, where we wrote the $g$-dependence of $c$ explicitly for the sake of clarity. Based on our observations in Fig.~\ref{kin-G} that $c(g)$ changed a few percent as $g$ was increased by 10\%, we assumed that the thermal velocity at $t>0$ is given by $c(g)$ if $\Delta g$ is sufficiently small. Thus, these time scales at $t>0$ have the form $\tau= c(g)/(g+\Delta g) \times \mbox{const.}$, where $g+\Delta g$ is the gravitational acceleration at $t>0$. This dependence of all the characteristic time scales on $\Delta g$ leads us to the scaling relation
\begin{eqnarray}
 \chi(\omega;g, g+\Delta g)=\frac{1}{M}\left(\frac{c(g)}{g+\Delta g}\right)^2 \hat{\chi}\left(\omega \frac{c(g)}{g+\Delta g}\right),
\label{scaling}
\end{eqnarray}
where $\hat{\chi}$ is a non-dimensional function.

As long as Eq.~(\ref{scaling}) holds, we can estimate the limit in Eq.~(\ref{chiomegag}) as
\begin{eqnarray}
 \chi(\omega;g)=\left(\frac{g+\Delta g}{g}\right)^2
\chi\left(\omega\frac{g+\Delta g}{g}; g, g+\Delta g\right).
\label{chi-scaled}
\end{eqnarray}

\begin{figure}[htbp]
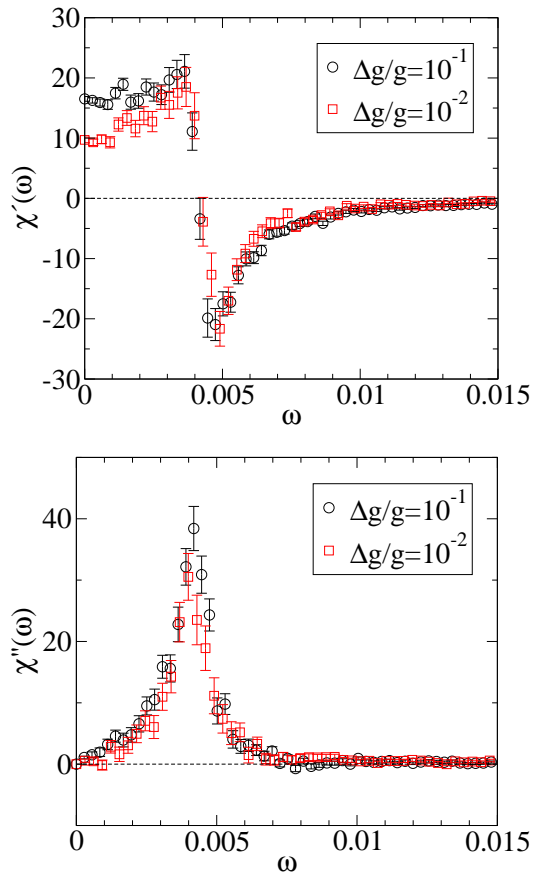

\begin{center}
\subfigure{%
\includegraphics[width=0.8\columnwidth,clip]{./fig8_top.eps}}%

\subfigure{%
\includegraphics[width=0.8\columnwidth,clip]{./fig8_bottom.eps}}%
\caption{\label{chi-r0992} (Color online)
The real part $\chi'$ (top) and the imaginary part $\chi''$ (bottom) of the frequency response function versus angular frequency $\omega$ for $r=0.992$. Circles show the data for $\Delta g/g=10^{-1}$ using Eq.~(\ref{chi-scaled}) with averages obtained over $800$ realizations. Squares are for $\Delta g/g=10^{-2}$ without using Eq.~(\ref{chi-scaled}); averages were obtained over $41000$ realizations.
}
\end{center}
\end{figure}

To verify the validity of the scaling relation Eq.~(\ref{scaling}), we performed two series of simulations for $r=0.992$. First, we measured the function $\chi(\omega; g,\,g+\Delta g)$ in Eq.~(\ref{chideltag}) for $\Delta g/g=10^{-2}$, taking the average over $41000$ realizations. Second, we measured the frequency response function $\chi(\omega;g)$ using Eq.~(\ref{chi-scaled}) for $\Delta g/g=10^{-1}$, taking the average over $800$ realizations. In Figs.~\ref{chi-r0992}, we compare the frequency response functions obtained from these two series of simulations. We found that they were consistent, although there were some discrepancies in $\chi'(\omega)$ near $\omega=0$.

More evidence of validity of the scaling relation comes from the fact that an FDR in an equilibrium system is satisfied when we measured the frequency response function using Eq.~(\ref{chi-scaled}). This is discussed further later (see Fig.~\ref{fdr-r1}).

\begin{figure}[htbp]
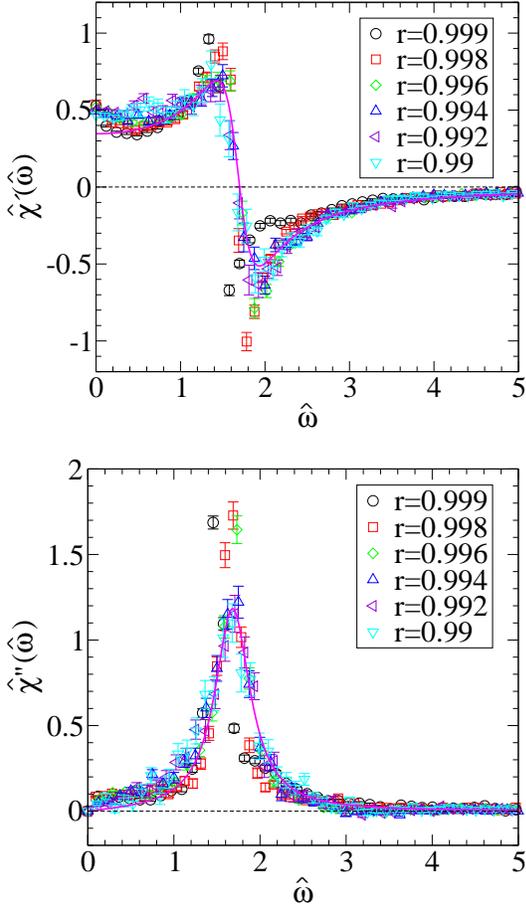

\begin{center}
\subfigure{
\includegraphics[width=0.8\columnwidth,clip]{./fig9_top.eps}}%

\subfigure{
\includegraphics[width=0.8\columnwidth,clip]{./fig9_bottom.eps}}%
\caption{\label{sc-chi} (Color online)
Real (top) and imaginary (bottom) parts of the scaled frequency response function versus the scaled angular frequency $\hat{\omega}=\omega c/g$. Averages were taken over 800 realizations. The thick lines are the theoretical prediction Eqs.~(\ref{scaledchireal}) and (\ref{scaledchiimag}) with fitting parameters $\hat{\mu}=0.50$, $\hat{\Omega}=1.7$.
}
\end{center}
\end{figure}

The frequency response functions presented below were obtained using the scaling relation Eq.~(\ref{scaling}) (and Eq.~(\ref{chi-scaled})) by averaging over $800$ realizations. In Figs.~\ref{sc-chi}, we show the real (top) and imaginary (bottom) parts of the scaled response functions $\hat{\chi}'$ and $\hat{\chi}''$ as functions of $\hat\omega$. Here, values of $c$ in Eq.~(\ref{scaledchireal}) and (\ref{scaledchiimag}) were calculated in an NESS without perturbation.
Theoretical predictions Eqs.~(\ref{scaledchireal}) and (\ref{scaledchiimag}) with the same (universal) fitting parameter as estimated in Fig.~6, $\hat{\mu}=0.50$ and $\hat{\Omega}=1.7$, are shown by thick lines. It appears that $\hat{\chi}(\omega)$ is consistent with the theoretical predictions if $r\le 0.996$.

\subsection{Fluctuation-dissipation relation}
To test the FDR Eq.~(\ref{fdt}) predicted by our theory, we evaluate the left- and right-hand sides independently using the results of simulations on $S(\omega)$ (Sec.~II~B) and $\chi''(\omega)$ (Sec.~II~C) presented in previous subsections. Note that $\overline{K}_{\rm COMz}$ were measured in the NESS where $S(\omega)$ was measured.

\begin{figure}[htbp]
\begin{center}
 \includegraphics[width=0.8\columnwidth,clip]{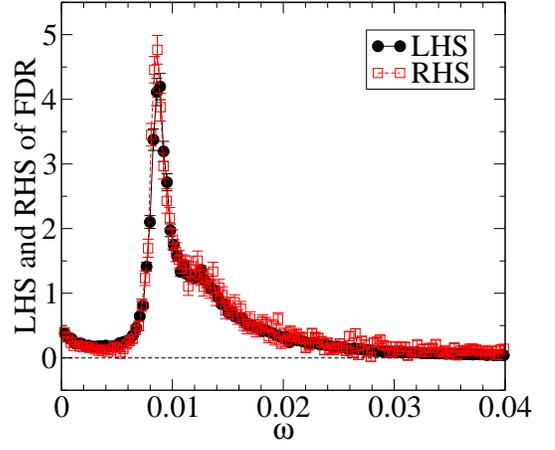}
\caption{\label{fdr-r1} (Color online)
Left-hand side $\omega S(\omega)/2k_B T_{eff}$ and right-hand side $\chi''(\omega)$ of Eq.~(\ref{fdt}) for $r=1$. $N=5000$, $L=100$, and $g=10^{-2}$, with averages over 400 realizations for $S(\omega)$ and over 800 realizations for $\chi''(\omega)$.
}
\end{center}
\end{figure}

\begin{figure}[htbp]
\begin{center}
 \includegraphics[width=1.0\columnwidth,clip]{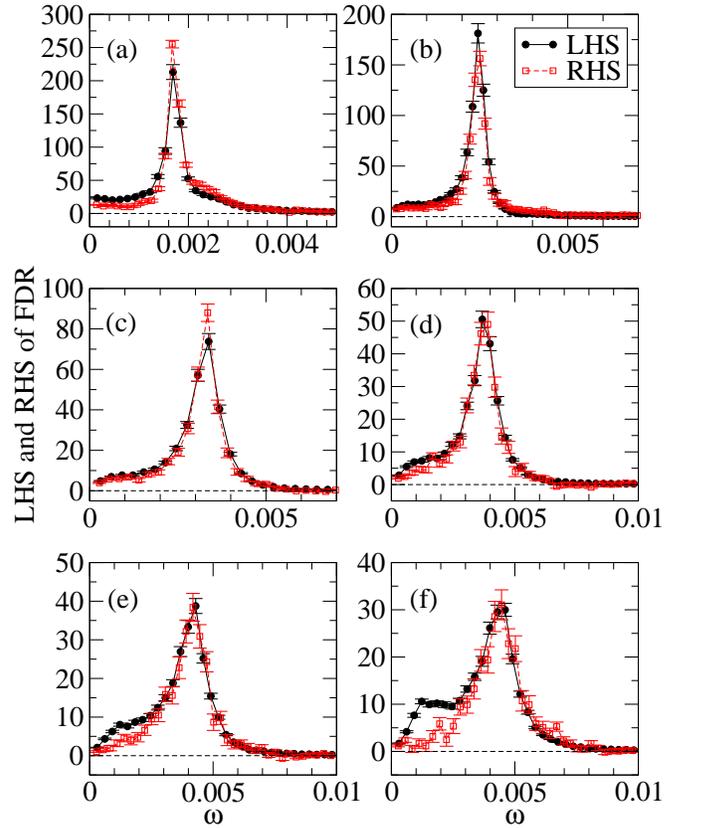}
\caption{\label{fdr-r0999-r099} (Color online)
Left-hand side $\omega S(\omega)/2k_B T_{eff}$ and right-hand side $\chi''(\omega)$ of Eq.~(\ref{fdt}) for (a) $r=0.999$, (b) $r=0.998$, (c) $r=0.996$, (d) $r=0.994$, (e) $r=0.992$, and (f) $r=0.99$. $N=5000$, $L=100$, and $g=10^{-3}$, with averages over 400 realizations for $S(\omega)$ and over 800 realizations for $\chi''(\omega)$.
}
\end{center}
\end{figure}

First, we confirmed that the FDR held within the error bounds of the simulation result in the whole range of $\omega$ given $r=1$ in Fig.~\ref{fdr-r1}. The stationary state is just the equilibrium state of elastic particles on a thermal wall. In Fig.~\ref{fdr-r0999-r099}, the left- and right-hand sides of the FDR are plotted as a function of $\omega$ for different $r$ values. For all $r$ ($0.99 \le r \le 0.999$), we found that the FDR held within the error bounds in the higher frequency range of $\omega$, including a region near the highest peak at $\omega=\omega_p$. The angular frequency of the highest peak $\omega_p$ was close to $\omega_{\rm osc}$, defined as the angular frequency of a peak in $S(\omega)$. We stress here that we defined $T_{eff}$ as $T_{eff}=2\overline{K}_{\rm COMz}/k_B$ in the FDR. The quantitative agreement in Fig.~\ref{fdr-r0999-r099} supports this definition of $T_{eff}$, using $\overline{K}_{\rm COMz}$ instead of using the global granular temperature $T$, because $T_{eff}$ is more than three times
larger than $T$ for
$r\le 0.996$ (see Fig.~\ref{kineticenergies}).

We found systematic deviations in the region $\omega < \omega_{p}$ for $r=0.999$ and $r\le 0.994$. These deviations were related to the fact that there was a peak near $\omega = 0$ in $S(\omega)$ (shown in Fig.~\ref{pscom}), while no corresponding peak near $\omega = 0$ appeared in $\chi''(\omega)$. As we discussed in Sec.~IV~B, the peak near $\omega= 0$ could have been connected with slow fluctuations of granular temperature due to thermal conduction and collisional dissipation. For $r=0.999$, the time scales of these two processes ($\tau_{\rm therm}$ and $\tau_{\rm diss}$) became much larger than $\tau_{\rm osc}$. Hence, the fluctuations with long lifetimes might be responsible for the deviations at small $\omega$. For $r\le 0.994$, the deviation appeared to increase as $r$ decreased. Because the system had lower granular temperature for smaller $r$, a larger heat current from the thermal wall was induced, causing larger fluctuations in global granular temperature. Further investigation is necessary to understand this
violation of the FDR more
precisely.

\section{Conclusion}
We studied the validity of the fluctuation-dissipation relation with regard to the COM motion in an NESS of a driven granular fluid under gravity. By neglecting the fluctuations of global temperature caused by thermal conduction and collisional dissipation, which change much slower than the macroscopic oscillation of the fluid, we derived a Langevin equation for the COM height using phenomenological considerations. This equation predicts functional forms of the correlation and response functions for the COM height that contain two phenomenological numerical constants $\hat{\mu}$ and $\hat{\Omega}$, which are used as fitting parameters. It also gives a fluctuation-dissipation relation accompanied by an effective temperature $T_{eff}$ that characterizes the agitating motion of the COM height by $T_{eff}=2\overline{K}_{\rm COMz}/k_B$.

To test the fluctuation-dissipation relation, we performed event-driven MD simulations and measured the power spectrum and response function for the COM height. While the power spectrum was consistent with our theory for $r\le 0.996$ and $\omega$ around the angular frequency of the slowest oscillation of the COM, it also showed large deviations from the theoretical predictions near $\omega=0$ for all $r$ ($0.99\le r\le 0.999$) and in the whole range of $\omega$ for $r> 0.996$.
The response function agreed closely with our theory for $r\le 0.996$ but showed deviations for $r> 0.996$. Furthermore, we compared the left- and right-hand sides of the FDR. The results showed that the FDR held in a region of $\omega$ near the highest peak for all cases of $r$ we tested. It was violated near $\omega=0$ for small $r$, $r\le 0.994$, and for $r$ close to unity, $r=0.999$. For $r\le 0.994$, the violation became more pronounced as $r$ decreased. The violation of the FDR was attributed to a peak near $\omega=0$ in the power spectrum for the COM height, which was absent in the imaginary
part of the frequency response function. The peak near $\omega=0$ in the power spectrum cannot be described by our theory.

We showed that these deviations near $\omega=0$ could be attributed to slow fluctuations of global temperature, defined as the slowly varying part of the kinetic energy per particle $K(t)$ due to thermal conduction and collisional dissipation. These fluctuations of global temperature were neglected in our theory. The deviations in the power spectrum and resulting violation of the FDR are expected to be accounted for by a theory that describes both $Z(t)$ and $K(t)$, which we will investigate in the future.

In Ref.~\cite{HaradaSasa-2005} a formula that connects the violation of the FDR in an NESS with the energy dissipation, or equivalently the energy input from outside, was proposed. A theory extended to include the slow dynamics of $K(t)$ and direct measurement of energy input in our simulations might give some insight into the generality of their formula.

Finally, the basic question of whether the effective temperature $T_{eff}$ obtained here has any physical meaning in terms of thermodynamics remains. The definition of effective temperature in a system that relaxes in several time scales, typically glass, has been debated in Refs.~\cite{CugliandoroKurchanPeliti-1997,BerthierBarrat-2002,Cugliandolo-2011}. It would be interesting to apply their theories to our problem with three time scales $\tau_{\rm therm}$, $\tau_{\rm diss}$, and $\tau_{\rm osc}$. It would also be interesting to investigate via simulation what happens if two systems with different effective temperatures are in contact with each other. Measuring the direction of heat flow directly might clarify the physical meaning of the effective temperature.

\begin{acknowledgments}
J. W. is grateful to H. Nakanishi, T. Sakaue, T. Saito, and C. Nakajima for their hospitality during his stay at Kyushu University, where part of this study was done.  This study was supported by the Grant-in-Aid for Scientific Research from the Ministry of Education, Culture, Sports, Science and Technology No. 23740293. Some of the computations for this study were performed using the facilities of the Supercomputer Center, the Institute for Solid State Physics, the University of Tokyo, and the Research Center for Computational Science (RCCS) in Okazaki, Japan.
\end{acknowledgments}

\appendix
\section{Langevin equation}
 In this section, we summarize the derivation~\cite{WakouOchiaiIsobe-2008,WakouIsobe-2010} of the Langevin equation that describes the motion of the COM of grains.

The equation of motion for the COM of the grains in the model described in Sec.~II can be written as
\begin{equation}
M\frac{d^2 Z}{dt^2}=-Mg+F_b.
\label{eqmotion}
\end{equation}
The right-hand side of the equation of motion for the COM must be, in general, the sum of the external forces acting on the grains. In our model, these are the gravitational force $-Mg$ and the $z$-component of the force exerted by the bottom wall $F_b$. Thus, it is essential to understand the properties of $F_b$ for the study of COM motion.

\begin{figure}[htbp]
\begin{center}
 \includegraphics[width=8.0cm]{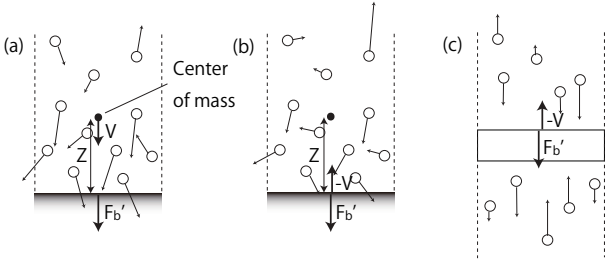}
\end{center}
\caption{\label{sketch}
(a) Schematic of the system observed in the laboratory frame of reference, (b) the same system observed in the center of mass frame, (c) the Rayleigh piston: a piston that undergoes random collisions with a one-dimensional heat bath of particles.
}
\end{figure}

Let us consider the reaction force $F'_b(=-F_b)$: the force exerted by grains against the bottom wall. A snapshot of the granular fluid is sketched in Fig.~\ref{sketch} (a). Suppose the COM is at a height $Z$ and is moving downward at velocity $V$. We now change the frame of reference to the center of mass frame (see Fig.~\ref{sketch} [b]); the bottom wall that lies a distance $Z$ away from the COM in the $z$-direction is moving upward with velocity $-V$. Now the problem is how to determine $F'_b$, the force acting on the bottom wall as a result of frequent collisions of granular particles, in the situation shown in Fig.~\ref{sketch} (b). There, the bottom wall is moving upward with velocity $-V$ against the macroscopically static fluid.

This problem is similar to the problem of determining the force acting on a one-dimensional Brownian particle (the Rayleigh piston~\cite{vanKampen}) moving in the $z$-direction with a velocity $-V$ (see Fig.~\ref{sketch} [c]). We create an expression for $F'_b$ on the basis of this analogy and assume that $F'_b$ consists of three components. The first is a systematic force $f_P(t)$ that equals the pressure multiplied by the area of the bottom wall. Because the local density near the bottom wall changes according to the motion of the COM, this force may depend on time. Apparently, the long time average of $f_P(t)$, that is, $\overline{f_P}$, must be equal to $-Mg$, the gravitational force acting on all particles. The simplest assumption for the time-dependent part of $f_P(t)$ is that it is proportional to the deviation of the COM height from its stationary value, $Z(t)-\overline{Z}$. This is because the change in local density near the bottom wall is proportional to $-(Z(t)-\overline{Z})$ if the change in the
height of the COM is sufficiently small: $|Z(t)-\overline{Z}|/\overline{Z}\ll 1$. The second component is a frictional force. We assume here the simplest form of the frictional force: linear in the relative velocity $-V(t)$ of the bottom wall to the COM. The third component is a random force. We assume
\begin{eqnarray}
 F'_b(t)&=&-Mg + M\Omega^2\left(Z(t)-\overline{Z}\right)+M\mu V(t)+R'(t)
\nonumber\\
&=&-F_{b}(t),
\label{fbdash}
\end{eqnarray}
where $\Omega$ is a coefficient that specifies the angular frequency of the slowest oscillation of the COM, and $\mu$ is the frictional coefficient. According to the discussion of characteristic time scales in Sec.~II~B, the time scales for macroscopic oscillation $\tau_{\rm osc}$ and that for pressure relaxation $\tau_{p}$ are $\tau_{\rm osc}\sim\tau_{p}\sim c/g$. Thus, we assume
\begin{eqnarray}
\Omega=\hat{\Omega}/\tau_{\rm osc}=\hat{\Omega}g/c,\hspace{0.5cm}
\mu=\hat{\mu}/\tau_{p}=\hat{\mu} g/c.
\end{eqnarray}

For the random force, we assume stationary Gaussian white noise in the same way as for the Rayleigh piston:
\begin{equation}
\left<R'(t)\right>=0,\quad \quad \left<R'(t)R'(t')\right>=I\delta(t-t').
\label{gaussianwhite2}
\end{equation}
where $I$ represents the intensity of the random force.

Substituting the $F_b$ obtained in (\ref{fbdash}) into the equation of motion of the COM (\ref{eqmotion}), we obtain
\begin{eqnarray}
\frac{d^2\delta Z}{dt^2}=-\Omega^2\delta{Z}-\mu\frac{d \delta Z}{dt} + \frac{R(t)}{M},
\label{langevin1}
\end{eqnarray}
where $\delta Z\equiv Z(t)-\overline{Z}$ and $R(t)=-R'(t)$. The random force $R(t)$ has exactly the same property described in (\ref{gaussianwhite2}) as $R'(t)$. Note that the Langevin equation (\ref{langevin1}) has the same form as that describing Brownian motion in a harmonic potential.

\section{Derivation of the power spectrum and the response function}
Derivation of the power spectrum and the response function from the Langevin equation describing Brownian motion in a harmonic potential is given in textbooks (see e.g., Ref.~\cite{ResiboisLeener}). We therefore present only essential steps in their calculation. First, we consider the power spectrum of the fluctuating motion of the COM obeying the Langevin equation (\ref{langevin1}). The formal solution of Eq.~(\ref{langevin1}) is written as
\begin{eqnarray}
Z(t)-\overline{Z}  =
\int_{-\infty}^{t}G(t-t')\frac{R(t')}{M}dt' +F_{ini}(t),
\label{cmheight}
\end{eqnarray}
where the function $G(t)$ is given by
\begin{equation}
G(t) = \frac{e^{-\frac{\mu}{2}t}}{\omega_0} \sin\left(\omega_0 t\right),
\label{gt}
\end{equation}
and $\omega_0$ is defined by $\omega_0\equiv(\Omega^2-(\mu/2)^2)^{1/2}$. The last term $F_{ini}(t)$ in Eq.~(\ref{cmheight}) consists of those that depend on the initial conditions and vanish after a sufficient amount of time. Thus, the term is negligible when calculating long time averages of physical quantities in the stationary state.

Using this formal solution, we can calculate the two-time correlation function $\phi(t)$ in an NESS defined by $\phi(t)\equiv\lim_{t'\to \infty} \left<\delta Z(t')\delta Z(t'+t)\right>$, where the brackets $\left< \cdots\right>$ indicate an average over the random force $R(t)$. We took the limit $t'\to \infty$ to ensure that the system is in the stationary state.

The power spectrum of $\delta Z(t)$ can be obtained using the Winner-Khinchin theorem:
\begin{eqnarray}
S(\omega)&=&\int^{\infty}_{-\infty} dt e^{-i\omega t} \phi (t)
\\
&=&
\frac{I}{M^2}\frac{1}{\left(\Omega^2-\omega^2\right)^2+\left(\mu \omega\right)^2}.
\end{eqnarray}

Next, we consider the response function for the COM, which describes the linear response of the COM with regard to a small external force $\varepsilon f(t)$. The Langevin equation in this case is written as
\begin{eqnarray}
\frac{d^2\delta Z}{dt^2}+\Omega^2\delta{Z}+\mu\frac{d \delta Z}{dt} - \frac{\varepsilon f(t)}{M} - \frac{R(t)}{M} =0.
\label{langevin2}
\end{eqnarray}
Taking the average over the random force, we obtain
\begin{eqnarray}
\frac{d^2\left<\delta Z\right>}{dt^2}+\Omega^2\left<\delta{Z}\right>+\mu\frac{d \left<\delta Z\right>}{dt}
- \frac{\varepsilon f(t)}{M} =0.
\label{langevin-average}
\end{eqnarray}

The response function $\chi (t)$ is defined as
\begin{eqnarray}
\left<\delta Z(t) \right>=\int^{t}_{-\infty} dt' \chi(t-t') \varepsilon f(t').
\label{linearresponse3}
\end{eqnarray}
Here, the external force $\varepsilon f(t)$ is assumed to be infinitely small. The Fourier transform of this relation yields $\langle \delta \tilde{Z}(\omega) \rangle=\chi(\omega) \varepsilon \tilde{f}(\omega)$, and hence
\begin{eqnarray}
\chi(\omega)=\lim_{\varepsilon\to 0}\left<\delta \tilde{Z}(\omega) \right>/ \varepsilon \tilde{f}(\omega).
\label{linearresponse2}
\end{eqnarray}

Performing the Fourier transform of the relation (\ref{langevin-average}) and comparing it with Eq.~(\ref{linearresponse2}), we obtain the frequency response function (complex admittance)
\begin{eqnarray}
\chi(\omega)=\frac{1}{M}\frac{1}{\Omega^2-\omega^2+i\mu\omega}.
\end{eqnarray}


\end{document}